\documentstyle[twocolumn,aps,prl,floats,epsf]{revtex}

\begin{document}


\twocolumn[\hsize\textwidth\columnwidth\hsize\csname %
@twocolumnfalse\endcsname

\draft

\title{Phonon Localization in One-Dimensional Quasiperiodic Chains}

\author{S.~E.~Burkov\cite{Burkov}, B.~E.~C.~Koltenbah, and
L.~W.~Bruch}


\address{Department of Physics, University of Wisconsin - Madison,
Madison, WI 53706}

\date{\today}

\maketitle
 

\begin{abstract}
Quasiperiodic long range order is intermediate between spatial
periodicity and disorder, and the excitations in 1D quasiperiodic
systems are believed to be transitional between extended and
localized. These ideas are tested with a numerical analysis of two
incommensurate 1D elastic chains: Frenkel-Kontorova (FK) and
Lennard-Jones (LJ). The ground state configurations and the
eigenfrequencies and eigenfunctions for harmonic excitations are
determined.  Aubry's {\it transition by breaking the analyticity} is
observed in the ground state of each model, but the  behavior of the
excitations is qualitatively different.  Phonon localization is
observed for some modes in the LJ chain on both sides of the
transition. The localization phenomenon apparently is decoupled from
the distribution of eigenfrequencies since the spectrum changes from
continuous to Cantor-set-like when the interaction parameters are
varied to cross the analyticity--breaking transition. The
eigenfunctions of the FK chain satisfy the ``quasi-Bloch" theorem
below the transition, but  not above it, while  only a subset of the
eigenfunctions of the LJ chain satisfy the theorem.
\end{abstract}

]

\narrowtext
 

There are surprisingly many examples of quasi-one-dimensional (1D)
quasiperiodic incommensurate systems, including quasicrystals, charge
density waves,  organic conductors, and various atomic monolayers
adsorbed on crystalline substrates. Even  high-$T_c$ superconductors
show incommensurate 1D-modulation. Although embedding these systems in
3D-space may change the outcome, understanding the conditions for
localization of excitations of the underlying model 1D-systems is an
essential beginning.
 
Both electron and phonon localization in 1D  incommensurate systems
have been studied rather extensively in recent years. Despite
considerable efforts and substantial success with several specific
models, the problem of formulating necessary and sufficient conditions
for localization is still far from being solved. The underlying
difficulty seems to be related to the specific position quasiperiodic
order takes among various spatial orderings: it is intermediate
between periodicity and disorder \cite{textbook}. On the one hand,
since quasiperiodic systems lack periodicity, conventional Bloch
theory does not apply. On the other hand, the degree of disorder in
quasiperiodic systems is not sufficient for Anderson's arguments about
localization in 1D disordered systems \cite{Anderson} to be applied to
the case. In fact Levitov showed\cite{Levitov} in a rather general way
that the eigenfunctions of the Schr\"odinger equation for a chain of
atoms change from extended to localized when the ``complexity" of the
atomic arrangements in the chain evolves from periodicity to
full-fledged disorder. Quasiperiodicity is the boundary between the
two regimes, and the eigenfunctions in a 1D  quasiperiodic system may
be  localized,  extended, or neither of the two. In the latter case
the eigenfunctions are called critical. As described in the following
paragraphs, examples of all three behaviors are found in various 1D
quasiperiodic systems. That is, the outcome is  model- and
system-dependent; even minor modifications to model Hamiltonians or
variations in parameters may  cause drastic changes in the
eigenfunction behavior. Thus,  characterizing the solutions of a 1D
Schr\"odinger equation with a quasiperiodic potential is a difficult
problem to solve analytically. In this paper we take a numerical
approach to demonstrating the dependence of localization on the model
and on the parameters. Since there is little difference in the formal
descriptions of electron and phonon localization, we study the latter,
because it is easier to adapt it to computer calculations without
losing essential mathematical details.

An example of extended wave functions in a 1D qua\-si\-pe\-ri\-od\-ic
system was first given in the Dinaburg-Sinai theorem  \cite{Dinsin}
for the differential Schr\"odinger equation
\begin{equation}
-\frac{d^2{\psi}}{dx^2} + V(x) {\psi} = E {\psi}  ,
\label{eq:1}
\end{equation}
where $V(x)$ is a weak analytic quasiperiodic potential.  The theorem
states that (i) the spectrum is continuous and (ii) the wave functions
are extended:
\begin{equation}
        {\psi}_k (x) = u_k (x) e^{ikx} ,         
\label{eq:2}
\end{equation}
where the Wannier functions $u_k (x)$ are quasiperiodic.
Unfortunately, mathematical constraints, mainly in the required
analyticity of $V(x)$, do not allow applying the theorem to many
systems of practical interest. Another  exactly solvable example is
the finite difference (discrete) Mathieu equation:
\begin{eqnarray}
& {\psi}_{n+1} - 2 {\psi}_n + {\psi}_{n-1}
+ V(n) {\psi}_n = E {\psi}_n, &
\label{eq:3} \\
& V(n) = V \cos ({\omega}n + {\alpha}), &             
\label{eq:4}
\end{eqnarray}          
with irrational ${\omega}$ and arbitrary phase $\alpha$. Aubry and
Andre \cite{Aubrycos} proved that for $V < 2$ almost all the wave
functions are extended and of the quasi-Bloch form Eq.(\ref{eq:2}).
These two examples  indicate that sometimes the Bloch theorem may be
applicable and the states be extended even when the Hamiltonian lacks
periodicity.
 
The discrete Mathieu equation, Eqs.(\ref{eq:3})  and (\ref{eq:4}),
also  gives an example of localization in quasiperiodic systems: for
$V > 2$ all wave functions are localized \cite{Aubrycos}, and the
spectrum is discrete:
\begin{equation}
{\psi}_n \propto e^{-n/l}, ~~~     n  \to {\infty} .                        
\label{eq:5}
\end{equation}
 
Critical behavior, i.e., neither localized nor extended, can be found
in 1D quasicrystals. They are usually modeled by the discrete
Schr\"odinger equation Eq.(3) with a quasiperiodic crystal potential
that takes only two values representative of two atomic species, A and
B, constituting the quasicrystal:
\begin{eqnarray}
        V(n) = \tilde V ({\omega}n + {\alpha}),\nonumber\\                      
        \tilde V(x+1)=\tilde V(x), \nonumber\\
        \tilde V(x) = V_A,  ~~~   \mbox{if} ~0<x<{\omega} \nonumber\\
                      V_B,    ~~~   \mbox{if} ~{\omega}<x<1 .
\label{eq:6}
\end{eqnarray}
This equation was extensively studied by many authors
\cite{Kadanoff,KKL,Ostlund} who gave rather convincing arguments of
critical behavior;  the mathematical proof was obtained by Bellisard
{\it et al}. \cite{Bellisard}. The spectrum is singular continuous, a
zero measure Cantor set of non-zero fractal dimension. The wave
functions are neither localized nor extended; moreover, they are even
divergent:
\begin{equation}
\int ~ [ |{\psi} (x)|^2 ]^2  ~ dx ~ \Big/ ~
\left[ \int ~ |{\psi} (x)|^2  ~ dx \right]^{2} ~ \propto  ~ L^{\gamma} ,  
\label{eq:7}                       
\end{equation}
where ${\gamma}$ is non-integer and eigenenergy-dependent and $L$ is
the length of the periodically repeated chain cells.
 
We believe that solving Eqs.(\ref{eq:3}) and (\ref{eq:6}) does not
constitute  a general solution for 1D quasicrystals.  Eq.(\ref{eq:6})
with two  undeformable atomic species, A and B, is too  crude a model
even for a 1D  quasicrystal chain.  Since the hard sphere
approximation leads to the potential $V(x)$ in Eq.(\ref{eq:6})  taking
only two values, one for each species, this in itself determines the
absence of localization for the model. It is a theorem that if a
quasiperiodic potential $V(x)$ takes only finite number of values then
localization is impossible \cite{DelyonPetritis}. To include the
possibility of localization, an elastic atomic chain should be
examined.
 

\begin{figure}[ht]
\begin{center}
\leavevmode
\epsffile{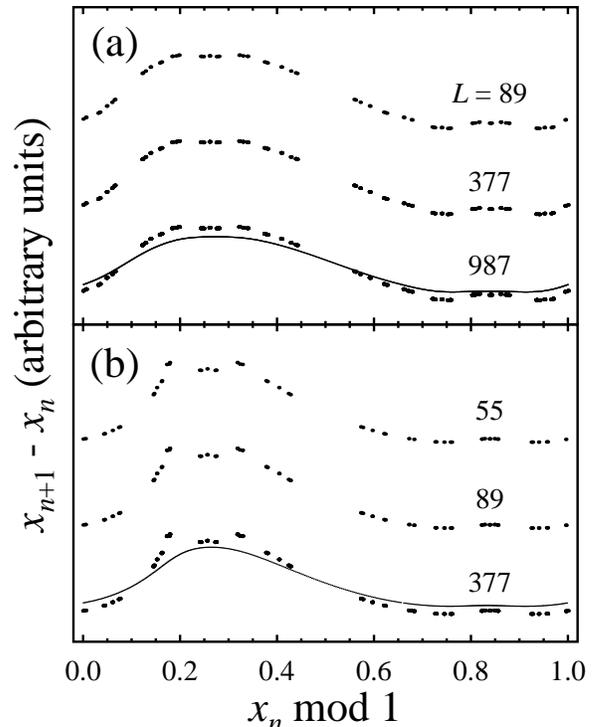}
\caption[]{
Phase space trajectories for the FK  and  LJ chains. The connected KAM
curves (small spots) and discontinuous  Cantorus curves (large spots)
are the trajectories  below and above  the transition by  breaking of
analyticity, respectively. The Cantorus results are shown for varying
chain sizes and are offset  from the KAM curve for clarity. The only
discernible difference, as a function of the cell length L, is  the
addition of spots at $x_n$ mod $1 = 0.67$ and $1.0$ from $L = 89$ to
$377$ in both models. (a) FK chain with  $V = 0.0207$ and $V = 0.0271$
for the small and large spots, respectively; (b) LJ chain with $V_g =
-80$ and $V_g = -110$, respectively, and with positions normalized to
$R_s = 3.25$ \AA.
}
\label{fig:traj}
\end{center}
\end{figure}

A simple model embodying many of the features of the above systems is
the 1D Frenkel-Kontorova (FK) chain:\cite{Pokrovsky-Talapov}
\begin{equation}
H =  \sum_{n} \left[\frac{1}{2} (x_{n+1} - x_n - a)^2
+ V \cos (2{\pi} x_n)\right].
\label{eq:8}
\end{equation}
Aubry and LeDaeron\cite{AubryFK} showed that the minimal  energy
configurations are periodic when the interatomic distance $a$ is
commensurate with the substrate period and  quasiperiodic when $a$ is
irrational. However, there are qualitatively different  configurations
for the two types of incommensurate states separated by  {\it the
transition by breaking of analyticity} predicted by Aubry. It is
customary to display this transition by plotting trajectories in the
phase space ($x_{n+1} - x_n$; $x_n$  mod $1$) as  in
Fig.~\ref{fig:traj}.  For each irrational $a$ there exists a critical
value $V_c$  of the substrate potential amplitude. If $V < V_c$,  the
trajectory derived from the atomic configuration $\{x_n\}$ is a smooth
analytic Kolmogorov-Arnold-Moser (KAM) curve; if $V > V_c$,  the
atomic configuration  $\{x_n\}$ is discontinuous, and the trajectory
is represented by a so-called Cantor set torus or Cantorus. The
critical value $V_c$ depends on $a$ in a rather complicated manner
\cite{AubryFK,BurkovSinai}. However, it is known that the largest $V_c
= 0.02461...$ is achieved when $a$ is equal to the Golden Mean
\cite{Greene}:
\begin{equation}
a = \frac{1}{2}(\sqrt{5} + 1) .   
\label{eq:9}
\end{equation}
We have restricted ourselves to this particular value of $a$ in the
numerical solutions.

The transition, despite its pure mathematical appearance, has profound
physical implications. The smooth KAM configurations keep a  vestige
of translational symmetry: the chain can slide without any change in
energy.  The discontinuous configurations are, on the other hand,
pinned by the substrate: there is a non-zero Peierls-Nabarro barrier
\cite{PeyrardAubry}. Thus, the transition by breaking of analyticity
coincides with an intrinsic pinning transition.  The spectrum of
harmonic excitations also changes there: in the KAM regime the  lowest 
eigenfrequency is zero (there is a translational Goldstone mode),
whereas in  the pinned Cantorus regime it  is non-zero
\cite{PeyrardAubry}. In view of these facts, one is tempted to
speculate that some  normal modes themselves evolve from extended to
localized when the system is driven over the transition by breaking of
analyticity. Our  numerical results show only a limited correlation
between the threshold for localization of the excitations and the
analyticity-breaking transition: the correlation does not hold for a
chain with anharmonic couplings and only a weak localization arises
for the harmonic FK chain, Eq.(\ref{eq:8}).
 
We checked two chains for phonon localization: the Frenkel-Kontorova
(FK) and the nearest-neighbor Lennard-Jones (LJ). The former is
described by Eq.(\ref{eq:8}), the latter by a similar Hamiltonian with
only a modification of the interatomic forces \cite{eq10}:
\begin{eqnarray}
H &=&  \sum_{n} [ J (x_{n+1} - x_n )
+ 2 V_g \cos (2{\pi} x_n/R_s) ], \nonumber\\  
        J(x) &=& 4 \epsilon [(\sigma/x)^{12} - (\sigma/x)^6]  .            
\label{eq:10}
\end{eqnarray}          
The results presented here are limited to the LJ chain with 
nearest-neighbor interactions, but tests with the interaction extended
to second and third neighbors showed no significant differences.
Critical values $V_c$ for the LJ chain depend on $\sigma$ as well as
on the mean misfit.
 
The first step of the solution was to determine minimum energy 
configurations for Fibonacci approximations to the golden-mean misfit
of both  chains\cite{GM} for various choices of parameters. For the FK
chain, a force--relaxation method was used, following guidance by
Aubry \cite{AubryFK,PeyrardAubry} on  how to avoid getting trapped in
local minima: for an initial  configuration with equidistant atoms, 
with the first atom in the ``proper" place, the system relaxes to the
absolute minimum. For the LJ chain, both the force relaxation and a
gradient search method were used to locate the minimum energy
configuration as there was an additional complication of possible
fracturing of the chain \cite{Miller,Milchev}. The transition by 
breaking of analyticity is clearly observable for both the FK  and LJ
chains, as shown in Fig.~\ref{fig:traj}. We show the smooth KAM curve
for the largest unit cells, L = 987 and 377 for the FK and LJ chains,
respectively.  The clumping of the Cantorus is shown there for several
values of L to demonstrate that its character is well-established at
modest values of L. 
 

\begin{figure}[t]
\begin{center}
\leavevmode
\epsffile{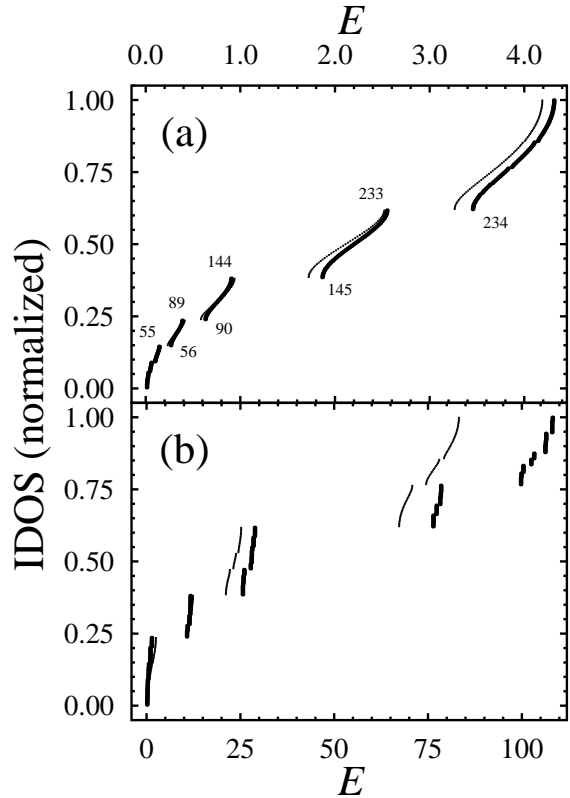}
\caption[]{
Integrated density of states for a chain size $L = 377$ for (a) the FK 
chain and (b) the LJ chain. In (a), the integers at the band edges are
the count of the eigenenergies  plotted  from left to right;  the
right-most numbers of each band are Fibonacci  numbers. This
coincidence with Fibonacci numbers also applies to (b). Note the
narrow width of the bands in (b) for the LJ model as compared with the 
bands in (a) the FK model. Potential parameters and labelings are as
in Fig.~\protect\ref{fig:traj}.
}
\label{fig:idos}
\end{center}
\end{figure}

In the second stage, tridiagonal dynamical matrices were constructed
for small--amplitude vibrations about the ground state configurations 
$\{x_n\}$ obtained in the first stage. This leads to the following
eigenvalue  problem for the FK chain for unit cell modes of zero
wavevector and atomic  mass $m$ 
\begin{eqnarray}
{\psi}_{n+1} - 2 {\psi}_n + {\psi}_{n-1} + V(n) {\psi}_n
=  - E {\psi}_n \nonumber   \\
V(n) =   (2 \pi)^2 V \cos(2 \pi x_n ),        ~~E = m{\omega}^2 ,
\label{eq:11}
\end{eqnarray}          
and to a similar result for the LJ chain, where the eigenenergies $E$
are related to the frequency by $E = m \omega^2 \sigma^2 / 4 \epsilon$
. Eqs.(\ref{eq:11}) are  an example of a general quasiperiodic
Schr\"odinger equation. However, unlike the case of Eq.(\ref{eq:6}),
the crystal potential $V(n)$ may take an infinite number of values.
Thus,  possible localization is not immediately evident
\cite{DelyonPetritis}. The eigenvalues and eigenvectors for the
dynamical matrices were obtained using standard ({\it EISPAC})
bisection routines.

\pagebreak
 

\begin{figure}[t]
\twocolumn[\hsize\textwidth\columnwidth\hsize\csname %
@twocolumnfalse\endcsname
\begin{center}
\leavevmode
\epsffile{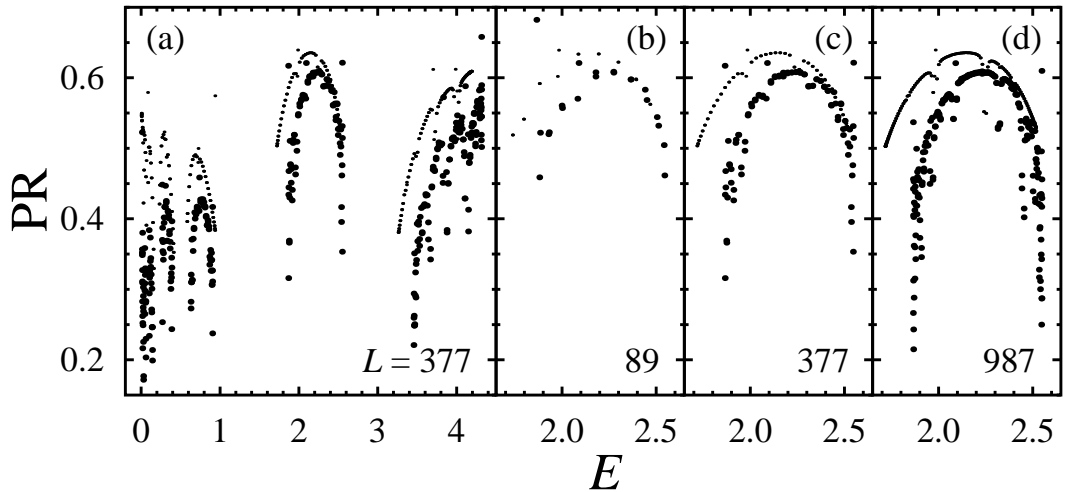}
\caption{
Participation ratio as a function of eigenenergy for the FK chain. (a)
full spectral range  for $L = 377$;  (b)--(d) over a selected range of
eigenenergies that corresponds to the central band of (a) for $L =
89$,  $377$, and $987$. Note the increase in the length of the
``tails'' of the disconnected Cantorus curves as the length of the
chain cell increases. Potential parameters  and labelings are as in 
Fig.~\protect\ref{fig:traj}(a).  
}
\label{fig:pr_fk}
\end{center}
]
\end{figure}

The minimal eigenvalue for both the LJ and FK chains changes  from
zero to non-zero when the system is driven over the transition by
breaking of analyticity in accordance with earlier observations by
Aubry \cite{PeyrardAubry}. For both models, the eigenvalue
distribution  also changes over the transition: at $V < V_c$, the
eigenvalues are evenly spaced, densely filling the bands as in the
periodic case; at $V > V_c$, however, the distribution is more
fragmented.   The resulting Integrated Density of States (IDOS) {\it
vs.} eigenenergy  $E$ (i.e., squared frequency ${\omega}^2$) and the
eigenvalue  distribution over the energy scale are qualitatively
similar  for the FK and LJ chains, as shown in Fig.~\ref{fig:idos}.
However, the apparent band widths for the LJ case are much narrower,
and surprising differences arise upon examining the spatial character
of the eigenfunctions. 
 
Before presenting the results for localization, we review some of the
limitations of a numerical study of the problem. Localization cannot
be decided by examining the eigenvectors of the dynamical matrix of
one rational approximation to the Golden Mean misfit. We followed the
evolution of results for a series of rational approximants converging
to the Golden Mean. Some properties of the eigenfunctions showed
strongly convergent behavior. There were also values of parameters and
parts of the eigenvalue spectrum where limitations of accuracy and
slow convergence combined to leave the situation undecided.  This is
the case of our rather weak evidence for  critical behavior in
localization.
 
Generally, it is difficult to draw quantitative conclusions about the
localization of eigenfunctions by visual inspection because of the
volume of information involved and because  oscillatory behaviour  is
sometimes mixed with the decay of a weakly localized state. The data
were first analysed for evidence of localization by computing
Participation Ratios ($PR$):
\begin{equation}
PR =  \frac{1}{L}
\frac{\left( \sum_{n}  {\psi}_n^2 \right)^2}{\sum_{n} {\psi}_n^4} .        
\label{eq:13}
\end{equation}
If the eigenfunction is extended, $PR$  tends to a finite limit as $L
\rightarrow \infty$.  If an eigenfunction is truly localized, as in
Eq.(\ref{eq:5}),  $L \times PR$ tends to a finite limit as the length
$L$ of the unit cell tends to infinity.  Critical states might have
more bizarre  scalings, as in Eq.(\ref{eq:7}) \cite{Zdetsis}.
 
\begin{figure}[t]
\begin{center}
\leavevmode
\epsffile{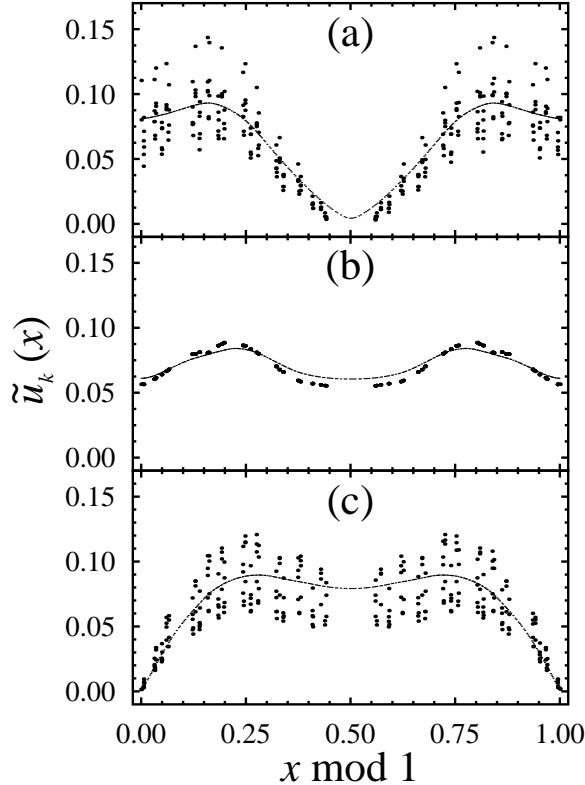}
\caption[]{
Quasi-Bloch functions $\tilde u_k(x)$ for the FK chain for $L = 377$ 
for three quasimomenta $k$ and configurations below and above the
transition, $V = 0.0207$ and $0.0271$, respectively. Parameters  and
labelings are as in Fig.~\protect\ref{fig:traj}(a). The values of $k$
\protect\cite{kvec} and the corresponding eigenenergies are  (a) $k =
1.95$, $E = 1.72$ and $1.87$; (b) $2.57$, $2.13$ and $2.22$; and (c)
$3.11$, $2.51$ and $2.55$.    The quasi-Bloch functions that are shown
correspond to states from (a) the left  tail (b) the middle peak and
(c) the right tail of Fig.~\protect\ref{fig:pr_fk}(c).  Note the
functions above the transition maintain the rough shape of the
corresponding functions below the transition but have a smearing that
is correlated  with the trend  of the participation ratios.
}
\label{fig:bloch_fk}
\end{center}
\end{figure}

We present the results of our localization studies on the
Frenkel-Kontorova chain first; participation ratios are shown in
Fig.~\ref{fig:pr_fk}. In the KAM  regime $V < V_c$, i.e., when the
ground state is smooth and unpinned, the $PR$ test shows that all
eigenfunctions are extended. This is corroborated by demonstrating
that the eigenfunctions can be transformed to the quasi-Bloch form
Eq.(\ref{eq:2}). More precisely, any quasiperiodic function (e.g., 
$u_{k} (n)$) may be represented as:
\begin{eqnarray}
        u_{k} (n) &=& \tilde u_{k} ({\omega}n + {\alpha}),
        \nonumber\\
        \tilde u_{k}(x)  &=& \tilde u_{k}(x+1) .
\label{eq:14}
\end{eqnarray}
The periodic generating function $\tilde u_{k}(x)$ in Eq.(\ref{eq:14}) 
is presented in Fig.~\ref{fig:bloch_fk} for three quasimomenta
\cite{kvec} that bracket the band shown in Fig~\ref{fig:pr_fk}(c).  At
$V < V_c$, the eigenvalues for  $L \rightarrow \infty$ become doubly
degenerate, as in the usual application of the Bloch theorem:   $E(k)
= E(-k)$. We determined the  value of the quasimomentum $k$ in       
Eq.(\ref{eq:2}) by matching  to the eigenfunctions of the  nearly
degenerate eigenvalue pairs. We were able to do this both below and
above the transition. Below, the function $\tilde u_k$ is  smooth.
Above, the eigenvalue pair typically differs in the fourth  decimal
place, but k-values could still be identified easily.  The function
$\tilde u_k$ above the transition has a clumping along the  abscissa,
as in Fig.~\ref{fig:traj}, and also in the ordinate. The departure
from the quasi-Bloch theorem becomes very pronounced at the band edge.
Above $V_c$, the participation ratios for states near the band edges  
decrease with increasing L, as shown in Fig.~\ref{fig:pr_fk}, but true
exponential localization was not observed. There may be critical
behavior, but more detailed analyses would be required to establish
that.
 

\begin{figure}[t]
\begin{center}
\leavevmode
\epsffile{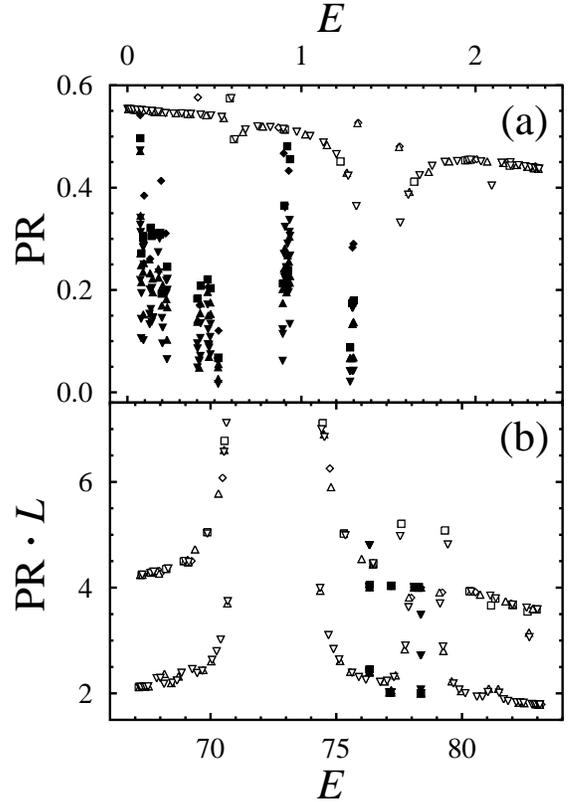}
\caption[]{
Participation ratio for the LJ chain over eigenenergies representative 
of the (a) ``extended'' and (b) ``localized'' regimes. Shown are
results for $L = 55$ (diamonds), $89$ (squares), $233$  (triangles)
and $377$ (inverted triangles). The open and filled symbols correspond
to $V_g = -80$ and $V_g = -110$, respectively, for the various chain
lengths (these are the same potentials  as in
Fig.~\protect\ref{fig:traj}(b)). The participation ratio is shown in
(a) whereas the participation ratio  times the chain length $L$ is
shown in (b) to emphasize the localization  over this range of
eigenenergies. Note the splitting of the curve in (b) as formed by the
renormalized  participation ratios for the various chain lengths.  
}
\label{fig:pr_lj}
\end{center}
\end{figure}

\begin{figure}[t]
\begin{center}
\leavevmode
\epsffile{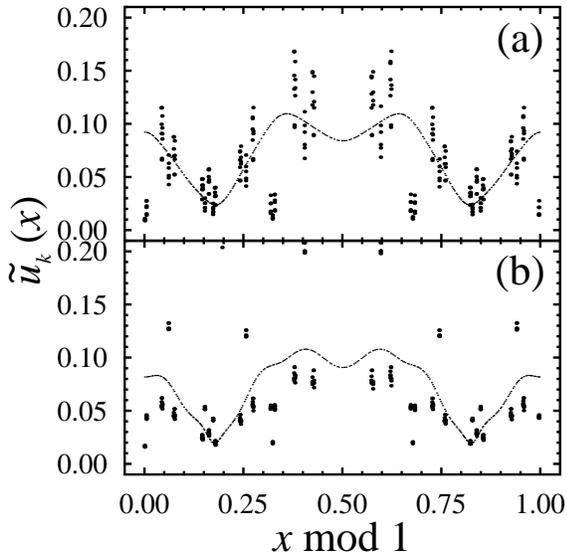}
\caption[]{
Quasi-Bloch functions $\tilde u_k(x)$ for the LJ chain for $L = 377$
at various  quasimomenta $k$, with labelings as in
Fig.~\protect\ref{fig:traj}(b). The values of $k$ and the
corresponding eigenenergies  are (a) $k = 0.38$, $E = 1.75$ and $0.91$ 
and (b) $0.42$, $2.03$  and $0.93$,  for $V_g = -80$ and $-110$,
respectively.  These values are  chosen from the  ``extended'' region
of Fig.~\protect\ref{fig:pr_lj}. Above the transition, the functions  
are smeared out, as in Fig.~\protect\ref{fig:bloch_fk} for the FK
chain, again correlated with the participation ratio.
}
\label{fig:bloch_lj}
\end{center}
\end{figure}

Participation ratios for part of the spectrum of the LJ chain  are
shown in Fig.~\ref{fig:pr_lj}. The behavior is very different from
that for the FK chain, even though the behavior of the eigenvalue
distribution  of the two chains is similar. There is a pronounced
high-frequency band of the LJ chain  where all eigenfunctions are
exponentially localized regardless of the smoothness of the underlying
ground state, i.e., localization occurs for both $V > V_c$ and $V <
V_c$. The localized upper band is by no measure small: the band
contains (Golden Mean)$^{-2} \sim 38\%$ of all states. The
construction  $L \times PR$ in Fig.~\ref{fig:pr_lj} converges with
successive Fibonacci  approximations to distinct values that might be
called localization lengths. In fact, and in contrast to the results
for the FK chain, examination of the corresponding eigenfunctions
shows  that the probability distributions $\psi_n^2$ are very narrow
and sharply peaked. They extend only a few atomic spacings  and their
participation  ratios  seem well-specified. The upper branch in 
Fig.~\ref{fig:pr_lj}(b) is precisely a factor of two larger than the
lower branch and corresponds to eigenfunctions that are concentrated
at two  spatial locations rather than one.  What we term the upper
band is itself a collection of smaller permitted and forbidden bands
with gaps at successively finer scales as the Golden Mean is more
closely approximated. The lower band  has more similarity to the FK
case, both in the  trends for the participation ratios and in the 
behavior of the quasi-Bloch functions, examples of which are shown in
Fig.~\ref{fig:bloch_lj}.  However, there are some differences between
the  quasi-Bloch functions of the two models: the LJ quasi-Bloch
functions shown for $V < V_{c}$ have more oscillatory features than
those of the FK functions,  and the clumping along the abcissa and
ordinate for $V > V_{c}$ is so much  more pronounced in the LJ chain
than in the FK chain that the remnant of the underlying functions can
barely be discerned. 


In conclusion, true, strong, phonon localization is observed in the
upper band of the nearest-neighbor LJ chain. In fact the observation
of a structure interpreted as a vibration within a domain wall for a
uniaxially modulated 2D lattice \cite{Gottlieb} was a motivation for
examining the 1D LJ chain. However, this localization is present
regardless of the changes in the level statistics and
pinning/unpinning of the chain, i.e., the transition by breaking of
analyticity. The behaviour of a closely related model, the FK chain,
is strikingly different: the transition by breaking of analyticity
does affect localization in the system. These results are in accord
with the widely held view that quasiperiodic chains lie on the border
between localization and delocalization and, thus, are very sensitive
to even minor perturbations. Not only is it inadequate to study
localization in quasiperiodic systems with models which correspond to
hard--core interactions, modest changes to compressible
nearest--neighbor interactions may have drastic effects on phonon
localization behaviour.
 

\section*{Acknowledgments}  This work was supported in part by the
National Science Foundation through Grants DMR-9120199, DMR-9423307,
and DMR-921473. 


\pagebreak


\begin{references}
\bibitem{eq10}  The notation for Eq. (9) is based on monolayer
calculations and the length and energy scales of the LJ potential are
chosen to be those previously used for xenon: $\epsilon = 230 K$ and
$\sigma = 4.0 \AA$. The results presented here are for $R_s = 3.25
\AA$, a value chosen so that  $(2 - a) R_s$ is close to the minimum of
the pair potential. Then the critical value of $V_g$ for the LJ chain
with mean spacing $(2 - a) R_s$ is approximately 100 K.
\bibitem{GM} The Golden Mean Eq.(8) was approximated by a converging
series of truncated continued fractions: 2/3, 3/5, 5/8,  ...
$F_n/F_{n+1}$,  ... , where $F_n$ are Fibonacci numbers: $F_{n+1} =
F_n + F_{n-1}$. The longest unit cell in this work corresponded to the
ratio 610/987, but most of the analyses were for cells  no larger than
233/377.
\bibitem{kvec} The wavenumbers were derived by fitting the
eigenfunctions  of nearly degenerate eigenvalue pairs, in the
KAM-regime, to Eq.(2). The  eigenfunctions in the Cantorus-regime were
associated with the corresponding eigenfunctions of the KAM-regime in
an enumeration in order of increasing  eigenvalue, and the wavenumber
of the KAM eigenfunction was used to construct an approximation to a
quasi-Bloch function. While this procedure has some arbitrariness, it
does seem to help systematize the data, as shown in
Fig.~\protect\ref{fig:bloch_fk}.
\end{references}
\end{document}